\documentclass[twoside]{article}
\usepackage{fleqn,espcrc2,epsf}

\newcommand{\beeq}{\begin{equation}}
\newcommand{\eneq}{\end{equation}}
\newcommand{\beeqa}{\begin{eqnarray*}}
\newcommand{\eneqa}{\end{eqnarray*}}

\newcommand{\bma}{\begin{displaymath}}
\newcommand{\ema}{\end{displaymath}}

\usepackage{graphicx}
\usepackage[figuresright]{rotating}

\newcommand{\AmS}{{\protect\the\textfont2
  A\kern-.1667em\lower.5ex\hbox{M}\kern-.125emS}}

\title{On the 1-loop lattice perturbation theory of the supersymmetric \\ 
         Ward identities.}

\author{Federico Farchioni\address{Deutsches Elektronen-Synchrotron, DESY, \\
        Notkestr.~85, D-22603 Hamburg, Germany},
         Alessandra Feo\address{Institut f\"ur Theoretische Physik,
                     Universit\"at M\"unster, \\ 
                     Wilhelm-Klemm-Str.~9, D-48149 M\"unster, Germany}
        \thanks{Talk given by Alessandra Feo},
        Tobias Galla\address{Department of Physics, University of Oxford,
                     Theoretical Physics, \\ 
                     1 Keble Road, Oxford OX1 3NP, UK},
        Claus Gebert$\rm^a$,
        Robert Kirchner$\rm^a$, \\
        Istv\'an Montvay$\rm^a$,
        Gernot M\"unster$\rm^b$
         \\[0.5em]
        DESY-M\"unster Collaboration \\[0.5em]}

\begin{document}

\begin{abstract}

The one loop corrections to the supersymmetric Ward identities (WIs)
in the discretized $N=1$ $SU(2)$ supersymmetric Yang-Mills theory 
can be investigated 
by means of lattice perturbation theory. The supersymmetry (SUSY) is
explicitly broken by the lattice discretization as well as by 
the introduction of Wilson fermions. 
However, the renormalization of the supercurrent can be carried out 
in a scheme that restores the nominal continuum WIs. 
We present our work in progress which is concerned with the 1-loop 
renormalization of the local supercurrent, i.e. with the perturbative 
computation of the corresponding renormalization constants and mixing 
coefficients. 
\vspace{1pc}
\end{abstract}

\maketitle
\section{INTRODUCTION}

The non-perturbative properties of SUSY gauge theories display many 
interesting features. For example, they may provide a possible 
mechanism for dynamical SUSY breaking.
If one believes in a fundamental theory of all interactions based on SUSY 
it is very important to construct realistic models incorporating 
supersymmetry.
The lattice regularization nowadays provides the only way to study 
non-perturbative effects from first principles.
SUSY is broken by the discretization of space-time itself due to the 
lack of lattice generators of the (continuous) Poincar\'e group.
Another problem is the question of how to balance bosonic and
fermionic modes, the numbers of which are constrained by the SUSY: 
the naive lattice fermion formulation produces too many fermions.

The simplest SUSY gauge theory is the $N=1$ SUSY Yang-Mills theory (SYM),
which is analogous to QCD with $N^2-1$ gluons and an equal number 
of Majorana fermions (gluinos) in the adjoint representation of
the colour group.
In the Wilson discretization of the SYM theory SUSY is
explicitly broken by the lattice itself and by the Wilson term, in
addition a soft breaking due to the gluino mass is present.
This formulation was originally proposed by 
Curci and Veneziano \cite{curci}.
SUSY is recovered in the continuum limit by tuning the bare parameters
(gauge coupling $g$, gluino mass $m_{\tilde{g}}$)
to the supersymmetric point.
The SUSY limit corresponds to $m_{\tilde{g}}=0$ 
and coincides with the chiral limit, in which the axial
symmetry is not explicitly broken to any order of perturbation theory.

In previous publications \cite{kirchner} we have investigated 
various aspects of the $N=1$ $SU(2)$ SYM theory. 
Recently, a different approach using domain wall fermions has been 
implemented \cite{fleming}.

The current interests of our collaboration focus on the study
of the SUSY WIs on the lattice, either from the numerical point of 
view (this is considered in another contribution to this conference
\cite{federico}) 
or by perturbatively calculating the renormalization constants and 
mixing coefficients of the local lattice supercurrent.


\subsection{SUSY Yang-Mills theory in the continuum}

In the continuum, the action for the $N=1$ SYM theory with an 
$SU(N_c)$ gauge group is
\beeqa
 && \mathcal{L} = -\frac{1}{4} F_{\mu\nu}^{a}(x) \, F_{\mu\nu}^{a}(x)      
                   + \frac{1}{2} \, \overline{\lambda}^{a}(x) \gamma_{\mu}
                   \mathcal{D}_{\mu} \lambda^{a}(x) \\
                   \nonumber\
                   & & \hspace{-1cm}+ (\textrm{gauge fixing} \,
                   + \; \textrm{ghost fields} \,
                   + \; \textrm{auxiliary fields})
\eneqa
where $\lambda^{a}$ is a 4-component Majorana spinor
and satisfies the charge conjugation condition
$\overline{\lambda}^{a} = {{\lambda}^{a}}^T C $.
The gluon fields are represented by 
$ A_\mu = -i g A_\mu^a T^a$ and  
$F_{\mu\nu} = -i g F_{\mu\nu}^{a} T^a$. 
${\cal D}_\mu\lambda^a=\partial_\mu\lambda^a+gf_{abc}A_\mu^b\lambda^c$ 
is the covariant derivative in the adjoint representation.

The continuum SUSY transformations read \\

\noindent
$\phantom{aaaaa} \delta A_\mu(x) = 
  - 2 g \bar \lambda(x) \gamma_\mu \varepsilon $,\\
$\phantom{aaaaa} \delta \lambda(x) = 
   -\frac{i}{g}\sigma_{\rho\tau} F_{\rho\tau}(x) \varepsilon $, \\
$\phantom{aaaaa} \delta \bar\lambda(x) = 
 \frac{i}{g} \bar\varepsilon \sigma_{\rho\tau}F_{\rho\tau}(x)$,\\

\noindent
where 
$\sigma_{\rho\tau} = \frac{i}{2} [\gamma_\rho,\gamma_\tau] $,
$\lambda = \lambda^{a} T^a $ and $\varepsilon $ is a global Grassmann 
parameter with Majorana properties.
These transformations relate fermions and bosons, leave the action 
invariant and commute with the gauge transformations so that the 
resulting Noether current $S_\mu(x)$ is gauge invariant. 
For the $N=1$ SUSY Yang Mills theory the supercurrent is a 
Rarita-Schwinger object \\
$ \phantom{aaaaa} S_\mu(x) = -2 \frac{i}{g} \mbox{Tr} \left\{ F_{\rho \tau}(x)
           \sigma_{\rho \tau} \gamma_\mu \lambda(x)  \right\}$. \\
Classically, the Noether current is conserved, 
$\partial_\mu S_\mu = 0$, provided the fields 
satisfy the equations of motion. Furthermore one finds that it 
fulfills a spin $3/2$ constraint: $\gamma_\mu S_\mu = 0$.

On the quantum level the nominal SUSY WIs are obtained by performing 
a variation of the 
functional integral with respect to a local transformation, 
$\varepsilon = \varepsilon(x)$. 
After putting the sources to zero we obtain
\begin{eqnarray*}&\big<{\cal O} \partial_\mu S_\mu(x) - 2 m_0 {\cal O} \chi(x) \big> + \\
&\big<\left.\frac{\delta {\cal O}}{\delta \overline{\varepsilon}(x)}\right|_{\varepsilon =0} - \left.
{\cal O} \frac{\delta S_{GF}}{\delta \overline{\varepsilon}(x)}\right|_{\varepsilon =0} - \left.
{\cal O} \frac{\delta S_{FP}}{\delta \overline{\varepsilon}(x)}\right|_{\varepsilon =0} 
\big>=0 ,
\end{eqnarray*}
where 
$\chi(x) = \frac{1}{2} F_{\mu \nu}^a(x) \sigma_{\mu \nu} 
\lambda^a(x)$. 
${\cal O}$ can be an arbitrary operator which is a function 
of the degrees of freedom of the theory (i.e. of $\lambda$ and $A_\mu$). 
The second line of this expression
corresponds to the contact terms (CT), the gauge fixing 
part of the action (GF) and the Fadeev-Popov term (FP), respectively.
This WIs is also discussed in \cite{dewit}.
\vspace{-0.1cm}

\subsection{Anomalies}

On the classical level
the axial current $ j_\mu^5$, the supercurrent
$S_\mu=-F_{\rho \tau}^a\sigma_{\rho \tau} \gamma_\mu \lambda^a$ 
and the energy-momentum tensor 
$\theta_{\mu \nu}$ belong to the same supermultiplet \cite{ferrara}, 
therefore they are related by a SUSY transformation.
The dilatation current is defined by 
$\hat{D}_\mu = x_\nu \theta_{\mu \nu} $, while the 
conformal spinor current reads ${\cal K}_\mu = x_\nu \gamma_\nu S_\mu $.
On the classical level the theory under investigation 
is scale and chiral invariant.
If quantum corrections are included anomalies 
for $j_\mu^5$ and $\hat{D}_\mu$ emerge.
The first one is the well known chiral anomaly,
which, for a suitable renormalization prescription, reads \\
$\phantom{aaaaa} < \partial_\mu j_\mu^5 > = 
- \frac{\beta(g)}{2 g} < F_{\mu \nu}^a \tilde{F}_{\mu \nu}^a> $, \\
and the second one is the trace anomaly \\
$ \phantom{aaaaa} < \partial_\mu \hat{D}_\mu >  \equiv <\theta_{\mu \mu}> =
\frac{\beta(g)}{2 g} < F_{\mu \nu}^a F_{\mu \nu}^a > $, \\
where $\beta(g)$ is the usual renormalization group $\beta$ 
function \cite{veneziano,jones}.
Similar equations (but with different coefficients) hold for QCD.

Because $j_\mu^5$, $\theta_{\mu \nu}$ and $S_\mu$ 
belong to a supermultiplet one would expect that\\
$\phantom{aaaaa} \partial_\mu {\cal K}_\mu = 
x_\nu \gamma_\nu \partial_\mu S_\mu  + \gamma_\mu S_\mu \ne 0 $, \\ 
so either $\partial_\mu S_\mu \ne 0$ or $\gamma_\mu S_\mu \ne 0 $.

In the early literature controversial results have been found. 
The matrix elements of the supercurrent have been calculated 
to 1-loop order using different regularization schemes. 
Some authors \cite{abbott} found
$ \partial_\mu S_\mu \neq 0$, 
while $ \gamma_\mu S_\mu =0$. Others 
\cite{hagiwara} found the divergence of the supercurrent 
to be zero but not its gamma-trace. 
Essentially these findings amount to either (1) or (2): \\
$ \phantom{aaaaa} \partial_\mu S_\mu =-2i\frac{\beta(g)}{g} \partial_\mu T_\mu
\, \, \mbox{and} \, \, \gamma_\mu S_\mu =0 \hfill (1)$ \\
\noindent
$ \phantom{aaaaa} \partial_\mu S_\mu = 0  \, \, \mbox{and} \, \,
\gamma_\mu S_\mu  = -4 \frac{\beta(g)}{g}\chi  \hfill (2)$ \\
where $T_\mu=-F_{\mu\nu}^a\gamma_\nu\lambda^a$. 
In case (1) $S_\mu$ is not conserved and has spin 3/2, 
while in case (2) $S_\mu$ is conserved, but has 
spin 3/2 $\oplus$ spin 1/2.

\subsection{Renormalization and mixing}

In a regularization, that breaks SUSY, the bare WIs reads, \\
$\phantom{aaaaa} \left< \partial_\mu S_\mu(x) \right> =  
2 m_0 \left< \chi(x) \right> + \left< X_S(x) \right> $, \\
where 
$X_S(x)$ is due to the breaking of the symmetry.
In order to renormalize the WIs, operator mixing has to be 
taken into account. 
$X_S(x)$ mixes with operators of equal or lower dimension. 
This includes $\partial_\mu T_\mu$, $ \partial_\mu S_\mu$ and $ \chi$.
With renormalization constants chosen appropriately one writes \\
$ \phantom{aaa} Z_S<\partial_\mu S_\mu(x)> + 
          Z_T<\partial_\mu T_\mu(x)> = $ \\
$ \phantom{aaaaa} 2 m_R Z_\chi<\chi(x)> + <\overline{X_S}(x)>$, \\
where $ < \overline{X_S}(x)> $ is forced to vanish in the continuum limit 
and $Z_S = 1 + O(g^2),\,Z_T = O(g^2)$.

Defining the renormalized supercurrent as \\
$ \phantom{aaaaa} S_\mu^R = Z_S S_\mu + Z_T T_\mu $, \\
one obtains 
$ \partial_\mu S_\mu^R  = 2 m_R Z_\chi \chi $.

In the chiral and SUSY limit $m_R = 0$.
Then we have $ \partial_\mu S_\mu^R  = 0$ and
$ \gamma_\mu S_\mu^R   = 2 i Z_T \chi $:  
this corresponds to case (2). 
Without taking into account the mixing 
(i.e. assuming $S_\mu^R = Z_S S_\mu$), one gets 
$ \partial_\mu S_\mu^R = -Z_T \partial_\mu T_\mu $ 
and $ \gamma_\mu S_\mu^R = 0$. 
Thus, the renormalization constants and the mixing coefficients 
can be fixed by imposing the nominal WIs to hold after renormalization.

\section{LATTICE FORMULATION}

Supersymmetric invariances cannot be carried over to formulations 
on a discrete space-time manifold due to the fact that it is 
not possible to fully adopt the (continuum) algebra of SUSY transformations.
In \cite{curci} Curci and Veneziano 
have used the standard Wilson formulation to discretize the $N=1$ 
SYM theory.
The gluonic part of the action is the standard plaquette action,
while the fermionic part of the action reads \\
$ S_f=\mbox{Tr}\Bigg\{ 
\frac{1}{2a}\Bigg(\bar\lambda(x)(\gamma_\mu-r)U_\mu^\dagger(x)
\lambda(x+a\hat\mu)U_\mu(x) $ \\
$ \phantom{aaaaa}-\bar\lambda(x+a\hat\mu)(\gamma_\mu+r)U_\mu(x)
\lambda(x)U_\mu^\dagger(x)\Bigg) $ \\
$ \phantom{aaaaaaaa} +\left(m_0+\frac{4r}{a}\right)\bar\lambda(x)
\lambda(x)\Bigg\}$.\\
\noindent
The lattice SUSY transformations can be chosen as follows
(${\cal G}_{\rho \tau}$ is the clover plaquette operator)\\

\noindent
$\delta U_\mu(x) = -
agU_\mu(x)\bar\varepsilon\gamma_\mu\lambda(x) \\
\phantom{aaaaaaaa}-ag\bar\varepsilon\gamma_\mu\lambda(x+a\hat\mu)U_\mu(x)$,\\
$\delta U_\mu^\dagger(x) =
+ag\bar\varepsilon\gamma_\mu\lambda(x)U_\mu^\dagger(x)  \\
\phantom{aaaaaaaa}+
agU_\mu^\dagger(x)\bar\varepsilon\gamma_\mu\lambda(x+a\hat\mu)$,\\
\hspace*{0.275em} $\delta\lambda(x) =
-\frac{i}{g}\sigma_{\rho\tau}{\cal G}_{\rho
\tau}(x)\varepsilon $,\\
\hspace*{0.275em} $\delta\bar\lambda(x) =
 \frac{i}{g}\bar\varepsilon\sigma_{\rho\tau}{\cal G}_{\rho
\tau}(x)$ \\

\noindent
and they reduce to the continuum SUSY transformations in the
limit 
$a \rightarrow 0$.

\subsection{SUSY Ward identities on the lattice}

On the lattice we may perform the same procedure as in the continuum 
to derive the SUSY WIs. 
Since the action is not fully supersymmetric, however, 
an additional breaking term $X_S$ appears in the bare lattice WIs.

Defining a local supercurrent on the lattice as\\
$ \phantom{aaaaa} S_\mu(x) = 
  -2 \frac{i}{g} \mbox{Tr} \left\{ {\cal G}_{\rho \tau}(x)
    \sigma_{\rho \tau} \gamma_\mu \lambda(x) \right\}$, \\
and taking into account all the contributions to the action 
we find for the bare WIs
\begin{eqnarray*}
&\big< {\cal O} \Delta_\mu S_{\mu}(x) -
2 m_0 {\cal O} \chi(x) - {\cal O} X_S(x)\big> \\
&+ \big<\left.\frac{\delta {\cal O}}{\delta \overline{\varepsilon}(x)}\right|_{\varepsilon =0} - \left.
{\cal O} \frac{\delta S_{GF}}{\delta \overline{\varepsilon}(x)}\right|_{\varepsilon =0} - \left.
{\cal O} \frac{\delta S_{FP}}{\delta \overline{\varepsilon}(x)}\right|_{\varepsilon =0} 
\big>=0 .
\end{eqnarray*}
$X_S$ is a complicated function of the link and the fermionic variables 
and its specific form depends on the choice of the lattice Noether current
(see for example \cite{taniguchi}). 
In the bare WI as formulated above $\Delta_\mu$ is to be understood 
as the symmetric lattice derivative.

In order to renormalize the WI on the lattice a possible operator mixing 
has to be taken into account.
$X_S$ mixes with operators of equal or lower dimension, 
$\Delta_\mu S_\mu $, 
$ \Delta_\mu T_\mu=-\Delta_\mu {\cal G}_{\mu\nu}^a \gamma_\nu \lambda^a $
and $ \chi = \frac{1}{2} {\cal G}_{\mu \nu}^a(x) \sigma_{\mu \nu} 
\lambda^a(x)$.

The standard way of renormalizing the supercurrent $S_\mu$ 
is to define a finite substracted $\overline{X_S}$ \\
$\phantom{aaaaa} \overline{X_S} = 
X_S+(Z_S-1)\Delta_\mu S_\mu+ 2 \overline{m} \chi+
Z_T\Delta_\mu T_\mu $, \\
and to impose that its expectation value fulfills  \\
$ \phantom{aaaaa} \lim_{a\to 0} <\overline{X_S}> = 0 $ \\
\cite{bochicchio}.
In this way we recover the renormalized WI
\begin{eqnarray*}
&\big< {\cal O} \Delta_\mu S_{\mu}^R(x) \big> = 
2 (m_0 - \overline{m}) Z_\chi^{-1}\left< {\cal O} \chi^R(x) \right> +  \\
& \big<-\left.\frac{\delta {\cal O}}{\delta \overline{\varepsilon}(x)}\right|_{\varepsilon =0} + \left.
{\cal O} \frac{\delta S_{GF}}{\delta \overline{\varepsilon}(x)}\right|_{\varepsilon =0} + \left.
{\cal O} \frac{\delta S_{FP}}{\delta \overline{\varepsilon}(x)}\right|_{\varepsilon =0} 
\big>.
\end{eqnarray*}
The CT, GF and FP terms of the action should also be renormalized.
We expect an additive renormalization due to the operator mixing:
this is related to the mixing with non gauge invariant operators 
as reported in \cite{taniguchi}.
The bare WI is an exact identity between matrix elements. 
By imposing directly that the renormalized WI has the nominal continuum form
\cite{dewit}, the calculation of $< X_S >$ can be avoided and traded
for the CT, GF and FP terms.
We have checked that the bare WI is fulfilled on tree level.

\subsection{State of the Art and Outlook}

\begin{figure}[htb]
\vspace*{-60pt}
\hspace*{-80pt}
\epsfxsize=13.0cm
\epsfbox{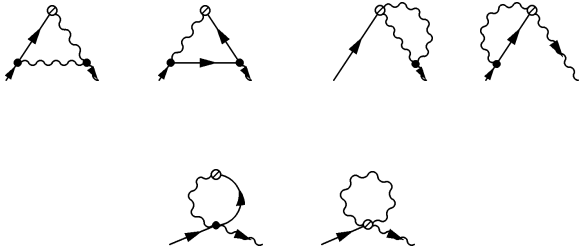}
\vspace*{-410pt}
\caption{One loop diagrams which contribute to the operator vertex correction
in the SUSY WI for the $S_\mu$ and $\chi$. }
\vspace*{-20pt}
\label{fig01}
\end{figure}

\begin{figure}[htb]
\vspace*{-30pt}
\hspace*{-80pt}
\epsfxsize=13.0cm
\epsfbox{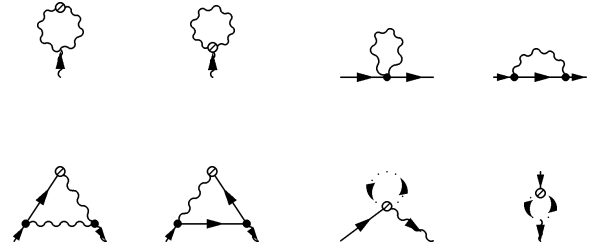}
\vspace*{-410pt}
\caption{One loop diagrams which contribute to the CT, GF and FP terms. }
\vspace*{-10pt}
\label{fig02}
\end{figure}

In our calculation we choose ${\cal O}:=A_\nu^b(y)\overline{\lambda}^a(z)$. 
The required correlations are given by 
$\left<{\cal O} \Delta_\mu S_\mu(x) \right>$, 
$\left<{\cal O} \chi(x) \right>$, corresponding 
to the diagrams in fig. \ref{fig01}, 
and  $\left< {\cal O} \frac{\delta S_{GF}}{\delta \overline{\varepsilon}(x)} |_{\varepsilon = 0} \right>$,
$\left< {\cal O} \frac{\delta S_{GF}}{\delta \overline{\varepsilon}(x)}
|_{\varepsilon = 0} \right>$ and 
$\left< {\cal O} \frac{\delta S_{FP}}{\delta \overline{\varepsilon}(x)}
|_{\varepsilon =0} \right>$,
corresponding to the diagrams in fig. \ref{fig02}. 

The computation of the matrix elements is carried out using the symbolic 
language of Mathematica. 
We have calculated the vertices for the composite operators
in fig. \ref{fig01},\ref{fig02} and nearly finished the calculation
of the contributions to the WI.
They are split into a logarithmically divergent part plus 
finite parts, the latter have been integrated by means of numerical 
routines.
We will present our results in a future publication.
It will be interesting to compare them with the numerical results 
\cite{federico}.

{\bf Acknowledgements:}
We would like to thank Gabriele Veneziano, Ken Konishi and 
Giancarlo Rossi for stimulating discussions.



\end{document}